\documentclass[a4paper]{article}
\usepackage{setspace} 
\usepackage{amsmath}
\usepackage[utf8]{inputenc} 
\usepackage{booktabs}
\usepackage{graphicx}
\usepackage[colorinlistoftodos]{todonotes}
\usepackage{bbm}
\usepackage{tikz}
\usetikzlibrary{intersections}
\usepgflibrary{arrows}
\usepackage{latexsym}
\usepackage{xcolor}
\usepackage{amsmath}
\usepackage{amsfonts}
\usepackage{graphicx}
\usepackage{eurosym}
\usepackage{multirow}
\usepackage{enumerate}
\usepackage{fancyhdr}
\usepackage{array}
\usepackage{colortbl}
\usepackage{mathabx}
\usepackage{arydshln}
\usepackage{hanging}
\usepackage[normalem]{ulem}
\usepackage{soul,xcolor}

\usepackage[margin=1in, footskip=0.25in]{geometry}

\definecolor{ballblue}{rgb}{0.13, 0.67, 0.8}

\begin{document}
\setstcolor{red}

\begin{center}
{\Large \bf  Sequential adaptive strategy for population-based sampling of a rare and clustered disease} 
\end{center}

{\large Fulvia Mecatti - {\em University of Milano-Bicocca}

\noindent (fulvia.mecatti@unimib.it, U7 Via Bicocca degli Arcimboldi 8, 20126 Milano, Italy)

\smallskip
Charalambos Sismanidis - {\em Global Tuberculosis Programme, WHO Geneva} 

Emanuela Furfaro - {\em Università Cattolica del Sacro Cuore, Milano}}

 \bigskip

\begin{center} {\bf \large Abstract} \end{center}

\noindent An innovative sampling strategy is proposed, which applies to large-scale population-based surveys targeting a rare trait that is unevenly spread over a geographical area of interest.  Our proposal is characterised by  the ability to tailor the data collection to specific features and challenges of the survey at hand. It is based on integrating an adaptive component into a sequential selection,  which aims to  both intensify detection of positive cases, upon exploiting the spatial clusterisation, and provide a flexible framework for managing logistical and budget constraints. To account for the selection bias,   a ready-to-implement weighting system is provided to release unbiased and accurate estimates.   Empirical evidence is illustrated from tuberculosis prevalence surveys, which are recommended in many countries and supported by the WHO as an emblematic example of  the need for an  improved sampling design.  Simulation results are also given to illustrate strengths and weaknesses of the proposed sampling strategy with respect to traditional cross-sectional sampling.


\bigskip

\doublespacing

\vspace{1cm}

\section{Introduction}
\label{sec:1}
This manuscript builds on the idea of improving the quality of sampled data by designing the sampling so that it is based on challenging features of the surveyed population. This idea can be traced back to the 1940s---with the Hansen-Hurwitz estimator (Hansen \& Hurwitz, 1943)---and the 1950s---with the Horvitz-Thompson estimator (Horvitz \& Thompson, 1952).  As early examples of inverse probability weighting, these estimators are unbiased, despite being based on data sampled with unequal selection probabilities for every  population units in the sampling frame. Since the 1980s, link-tracing designs have appeared in the literature and they have been used to deal with emerging issues, such as when the object of the survey is a rare population trait that is possibly unevenly spread over an area of interest. These sampling strategies now form a large class that are known as {\em Adaptive Sampling} (Seber \& Salehi, 2012). 
During the 25th Morris Hansen Lecture, S.K. Thompson  (2017)  recognised that {\em ``[...] an increasing attention on using sampling designs to make interventions effective in populations``}. An inspirational example for the need of such an innovative sampling design can be found in population-based surveys that measure tuberculosis (TB) prevalence at a national level, which are recommended in many settings around the world and are supported by the World Health Organisation (WHO) and partner agencies ({\em https://www.who.int/tb/challenges/task$\_$force/en/}). Despite being a global public health priority  (WHO, 2018), statistically speaking, TB prevalence among the general population qualifies as a {\em rare} trait that is measuring at less than 1\%, even in countries that are considered to have a high burden of the disease. Consequently, under a traditional self-weighting sampling approach, such as the one currently suggested in the most recent WHO guidelines, very large sample sizes (often around 100,000 people) are required for reasonable levels of estimation accuracy (Floyd {\em et al.}, 2011; Chapter 5). Consequently, the associated costs for these surveys are very high and are often a natural constraint for survey design. Survey costs are also inflated by measurement requirements, such as chest X-rays and laboratory tests on sputum specimens, to identify people with the disease.

Building upon the TB prevalence survey example, the main objective of this paper is the development of an innovative sampling strategy that aims to fully optimise the effectiveness of resources dedicated to the survey while, at the same time,   preserving and possibly improving the efficiency of the estimation. Our proposal is based on tailoring the data collection by combining an adaptive approach with a sequential selection. The adaptive component aims to purposely increase the detection of people with the disease by exploiting the spatial clustering that is typical of an infectious disease such as TB. An obvious benefit of this approach is its increased potential to make an impact in reducing TB disease burden upon the surveyed population given that once detected and put on appropriate treatment, most people with TB are cured. From an epidemiological perspective, finding and treating more cases of a rare disease allows for a better understanding of its epidemiology in a country, which in turn results in better informed public health action to control the disease. The sequential component of the sampling strategy proposed in this paper aims to provide a flexible framework for dealing with budget and logistic constraints at the design level of the survey, while at the same time fostering the goal of oversampling people with the disease. 

This paper is organised as follows. Section 2 provides the background for the methodology and the motivational example of TB disease. Section 3 introduces the starting point of a list-sequential adaptive strategy based on Poisson sampling, and derives the unbiasedness of resulting estimates and also their variance estimation. Section 4 discusses practical issues such as control over the sample size, which is typically random under an adaptive design. Section 5 discusses empirical evidence from previous surveys and  presents  simulation results  to highlight the strengths, weaknesses and areas for improvement over traditional sampling strategies. Section 6 outlines our concluding remarks and makes several recommendations for future research.

\bigskip \bigskip
\section{Background and Motivation}
\label{sec:2}
The term {\em adaptive} appears often in sampling statistics research. It is used both in survey design, which is referred to as {\em adaptive designs} (see e.g.  JOS Special Issue on Adaptive Designs, 2017), as well as in experimental design, with a fast-emerging literature on {\em adaptive} \ and {\em responsive} clinical trials (Tourangeau {\em et al.}, 2017). Link-tracing designs, which are the earliest example of adaptive design in the literature, were originally developed to deal with hard-to-sample/reach populations (Kalton \& Anderson,1986). They are characterised by an emphasis on the production of design-unbiased estimates using data that are collected {\em adaptively}. They now form a broad class of sampling strategies known as {\em Adaptive Sampling}. Popular examples of adaptive sampling include Network Sampling (Sirken, 2004), Adaptive Cluster Sampling (Thompson, 1990) and the recent addition Adaptive Web Sampling (Thompson, 2006). 

S. K. Thompson, who was doubtless  the main contributor to this class of sampling strategies, recently  stated (Thompson, 2017)
$<< [\dots]$ {\em 
an adaptive sampling design is one in which the selection of units to include in the sample depends on values of the variable of interest observed during the survey.} $>>$. He also suggested a third use
in addition to the two classical uses of a sampling strategy (i.e. to make inferences about population quantities and to set experiments onto populations), which is to apply interventions to impact changes in a population. We present an inspiring example of this latter use of sampling strategy:  the population-based surveys for assessing TB prevalence at a national level, promoted by WHO and its partner agencies in certain settings around the world ({\em https://www.who.int/tb/areas-of-work/monitoring-evaluation/impact\_measurement\_taskforce/en/}).

Worldwide, TB is one of the top 10 causes of death and it is the leading cause of death from a single infectious agent. In 2017, TB caused an estimated 1.3 million deaths among HIV-negative people and an additional 300,000 deaths among HIV-positive people, while 10 million people developed TB disease. TB remains a global priority in public health, particularly in Africa, South-East Asia and the West Pacific regions. The United Nations Sustainable Development Goals and the WHO’s End TB Strategy goals and targets provide the framework for national and international efforts to end the TB epidemic during the period 2016--2030. Monitoring  progress against epidemiological targets is possible using evidence from national surveillance systems with strong quality and coverage, complemented by periodic surveys and studies, particularly in settings where surveillance systems are still being strengthened. Perhaps the most important of periodic surveys are population-based surveys to measure the prevalence of TB disease. 

Currently, TB prevalence surveys are implemented according to the most recent international WHO guidelines (WHO, 2011). The recommended sampling design is a traditional, multi-stage, cross-sectional design, which is intended for general consumption by a wide array of practical users. Traditional sampling designs are relatively simple to implement and oblige  familiar notions, such as {\em representativeness},  that are widely understood among non-statistical audiences. Meanwhile, and more evidently in the case of TB prevalence surveys, traditional sampling designs have  limitations and inconveniences.

As already mentioned  TB is a rare population trait. Even in countries considered to have a high burden of TB disease, the level of TB prevalence is generally estimated to be less than 1\% among the national population at a given point in time. Consequently, under the currently recommended sampling design, this leads to very large sample sizes of  between 50,000--100,000 people to reach  a relative  precision  of the final estimate of 20\%--25\%, with an associated cost of USD 1--4 million. 

Furthermore, because TB is an infectious disease, people with TB are often clustered, i.e.  unevenly spread over the country, due to various epidemiological factors (e.g. demographic, socio-economic and cultural factors) and also due to health system factors (e.g. access to diagnostic and treatment services).  The main statistical consequence of such a spatial pattern for distribution of TB cases over the country is a usually large variability of TB prevalence between  areas across the country.

In the current WHO guidelines this between-area variability is accounted for in the standard formula for sample size determination (see WHO, 2011, Chapter 5,  formula (5.4) and the design effect correction), with the effect of  inflating the required sample size and further increasing the survey costs. This significant investment typically leads to the estimation of a national percentage figure based on the detection of a few people with TB among a very large sample of people without the disease.  An example of such outcome is given by the recent national TB survey conducted in Kenya in 2015--2016 according to the WHO guideline (Ministry of Health, Republic of Kenia, 2016) where 305  (bacteriologically confirmed) TB cases have been found  among the  63,050 participants in the survey  (see Section \ref{sec:5} for further discussion).

There is, of course, nothing methodologically wrong with the currently recommended traditional sampling design. However, the main stimulus behind the proposals presented in this manuscript is the potential for methodological improvements to optimise the investment of resources and effort, while at the same time to generate additional information for TB epidemiology in the setting in which the survey is implemented.

Another consideration for the development of a new sampling approach has been the drive to find more people with TB and, because TB is mostly treatable, cure them. An emphasis is given to over-sampling people with TB and making the survey itself a tool for reducing disease burden, generating new knowledge about TB epidemiology and informing public health action. This over-sampling has the potential to make the investment more effective and the survey itself a tool for impacting change on the general population, as theorised by S.K. Thompson (see Section \ref{sec:1}. Meanwhile, the over-sampling will purposely introduce a selection bias, which will need to be addressed at the estimation stage. Therefore, Adaptive Cluster Sampling appears as a natural candidate to enhance case detection by exploiting the spatial pattern rather than correcting for it. Sampling efforts will be concentrated in areas where TB prevalence is expected to be high, for instance higher than the national prevalence or higher than a chosen threshold.

Lastly, better control over logistics at the design level of the survey would be a desirable characteristic of the new strategy. For instance, being able to avoid logistically difficult areas of the country, areas that are hard to reach due to seasonal weather, flooding or even war zones. In these areas, data collection is typically compromised and the field operations budget is consequently increased. To this end, a {\em sequential selection} seems the right fit and it will be further explained in Section \ref{sec:3}.

Based on these considerations  we have  pointed out five key characteristics and properties that we intend to satisfy with our proposal for an improved sampling strategy. The following sections present an innovative sampling approach which builds upon these key design principles: 
\begin{enumerate}
\item Applies to surveying a rare population trait. 
\item Deals with geo-clustered population units, due to the existing spatial pattern leading to intra-cluster variability. The purpose here is to exploit the spatial pattern instead of correcting for it.
\item Leads to purposeful oversampling of units with the trait (e.g. people with TB disease) by  introducing selection bias during data collection. Therefore, the final sample is not {\em representative} of the target population, which needs to be corrected at the estimation stage by a proper weighting-system that addresses such {\em controlled} selection bias. 
\item Provides a flexible framework for dealing with logistical and implementation constraints.
\item Is feasible and statistically {\em simple} for use in general guidelines and field implementation.
\end{enumerate}

Despite using population-based TB prevalence surveys  as our inspirational example, we believe that this proposal could in fact be a useful blueprint for surveys on other  conditions, diseases, or population attributes.

\bigskip \bigskip
\section{Poisson Sequential Adaptive: sampling design and estimation}
\label{sec:3}
The starting point of our proposed strategy is the Poisson Sequential Adaptive (PoSA) sampling. We will focus on a finite population $U$ of $N$ individuals and on a binary survey variable $y$ taking the value $1$ if the person has TB disease and $0$ otherwise. Notice that these are both simplifications specific to the TB example. First  the sampling design currently suggested by WHO guidelines refers to a selection of areas and sub-areas; i.e., groups of individuals purposively informed according to specific suggested criteria. See the remark at the end of this section for a discussion about how PoSA may apply to selection of sub-areas (as primary sampling units) instead of direct selection of individuals. Second, defining and diagnosing TB is not a trivial task,  there are four  dedicated chapters in the WHO guidelines book which are devoted to case definition and diagnosis (WHO, 2011, Chapters 4 and 6 to 8) and further discussion is given in Section \ref{sec:5}.

Other components of the simplified setup that we consider are: {\em (a)} the choice of a non-enumerative sampling design (i.e. a design that would not require the listing of all possible samples, instead allowing unbiased estimation in a simple) and an analytic non-numerical closed form (e.g. the familiar Horvitz-Thompson (HT) estimator); {\em (b)} the simplest choice for an adaptive rule, which is still able to exploit the spatial clustering by intensifying the sample selection {\em close} to previously selected positive cases; and {\em (c)} the simplest choice for the spatial setting, namely reduced to one dimension.
In terms of sampling design, this last simplification would imply: 

\begin{itemize}
\item[\em(i)] The target population is pre-ordered according to a chosen convenient rule. As a result, population units would be either {\em close} or {\em distant} to each other according to such ordering. In the following, the labelling $U = \left\{1 \dots i \dots N\right\}$ is meant to reflect the chosen order. For instance   units $i-1$ and  $i$,    and units $i$ and $i+1$ are close for being subsequent, while units $i-1$ and $i+1$ are not.
\item[\em (ii)] The choice of a {\em List-sequential} sampling algorithm, which would follow the chosen order: all units included in $U$ are considered successively and at each step $i$ of the sequence $1 \cdots N$, a real-time decision is made whether unit $i$ should or should not be selected in the sample (see for example (Till$\acute{\text e}$, 2006, Chapter 3). 
\end{itemize}

In terms of our tuberculosis example, we can consider the geographical spatial pattern on the basis of TB being an infectious disease, so the one-dimensional simplification described above may be (for instance) a pre-designed route on a geographical map that is to be followed across the country by the field team. The choice of the route would be part of the design of the survey and could be tailored upon specific requirements and/or physical features of the country, as well as negotiated with local authorities. For instance, the route can be defined by minimising travel costs while at the same time acknowledging the presence of limited access areas. 

\smallskip
Notice that a list-sequential selection is in fact a standard basic sampling algorithm, which in theory allows the actual implementation of any sampling design. Bondesson and Thornburn (2008)  formalised a list-sequential methodology to produce a sample under very general conditions: {\em (i)} without replacement (WOR) to provide a final sample of all distinct units; {\em (ii)} assigning a generic inclusion probability to each population unit, that is not necessarily equal; and {\em (iii)} applying to either a fixed or a random sample size. In addition, this method, in the form of a ready-to-implement algorithm, is suitable for real-time sampling (as is the case of our TB example). We now illustrate the main tool of the Bondesson-Thornburn algorithm, which is an {\em updating matrix} that is able to formalise the entire selection process, while at the same time providing all of the analytic input required for unbiased estimation. 

The updating matrix lists the $N$ steps of the selection process on the rows, with $0$ as the {\em initial} state, and lists population units on the column in the chosen order, which also gives the visit/selection sequence. Let $\pi_i$ be the probability assigned to population unit $i$ to be selected into the sample, either the same for all units (equal probability design) or otherwise (unequal probability design). Let $S_i$ define the {\em Sample Membership Indicator} (SMI) of population unit $i$; i.e., a random variable taking value $s_i$ equal to $1$ if unit $i$ is selected into the sample and equal to $0$ otherwise. A random sample is then defined as the (random) vector of the $N$ SMIs, whose value $s_1 \dots s_i \dots s_N$, at the end of the sequential process, will indicate the selected sample. Starting from a given set of {\em initial} probabilities $\pi_i^{(0)}$, the updating matrix is given by

\begin{equation}
\label{updating matrix:def}
\begin{array}{r|cccccc} \small
unit \rightarrow & 1 & 2 & \cdots & i & \cdots & N \\
step \downarrow & & & & & & \\
\hline

1 & S_1=s_1 & \pi_2^{(1)} & \dots & \pi_i^{(1)} & \dots & \pi_N^{(1)} \\ \\
2 & s_1 & S_2=s_2 & \dots & \pi_i^{(2)} & \dots & \pi_N^{(2)} \\
\vdots & \dots & & \vdots & & & \vdots \\
i & s_1 & s_2 & \dots & S_i=s_i & \dots & \pi_N^{(i)} \\
\vdots & \dots & & \vdots & & & \vdots \\
N & s_1 & s_2 & \dots & s_i & \dots & S_n=s_N \\ \\
\end{array}
\end{equation}

\bigskip
The entries of the updating matrix show the current step-by-step state of the sampling process.
Notice that unit $i$ can be selected/not selected exclusively when {\em visited} at the $i$-th step of the sampling algorithm. Therefore, at each step of the selection sequence, $S_i$ can be in one of the three following states:
\begin{itemize}
\item[-] {\em Before} visiting unit $i$ (i.e. until step $i-1$), the sample membership of unit $i$ is a random event, as formalised through the random variable $S_i$;
\item[-] {\em At} step $i$ unit $i$ is {\em visited} and its selection/not selection attained (i.e. the sample membership indicator takes on its realisation $S_i=s_i$); and,
\item[-] {\em After} (i.e. from step $i+1$ on) the actual sample membership $s_i$ is recorded for unit $i$ with no more randomness.
\end{itemize}
Moreover, upon recording $S_i=s_i$ along the diagonal of the updating matrix, the selection probabilities of all subsequent (to be visited) units $j=i+1 \dots N$ are updated. Consequently, at the $i$-th step, unit $i$ is selected with updated probability $P\left(S_i=1\right)=E\left(S_i\right)=\pi_i^{(i-1)}$ located at the previous row, same column in \ref{updating matrix:def}.
Finally, the last row shows the selected sample. 

Bondesson and Thornburn (2008)
completed their list-sequential methodology by a suitable {\em updating rule}, given as a linear function of {\em updating weights} $w_{j-i}^{(i)}$ at each step $i$ for every unit $j \ge i+1$
\begin{equation}
\label{updating weights:def}
\pi_j^{(i)}=\pi_{j}^{(i-1)} - \left[s_i - \pi_i^{(i-1)}\right] w_{j-i}^{(i)}
\end{equation}
and showed that different properties of the sampling design are attained by a suitable choice for the updating weights. 
For instance, a final sample of fixed-size $n$ is given by setting $n=\sum_{i=1}^N \pi_i^{(0)}$ and by choosing updating weights with  unitary sum $\sum_{j=i+1}^N w_{j-i}^{(i)}=1$ row-by-row.

\smallskip
Our PoSA sampling proposal is based on Poisson sampling, which is the simplest list-sequential algorithm. This is a classic sampling design going back to the 1960s (H$\acute{\text{a}}$jek, 1964), which was  originally named {\em rejective}. Its simplicity is based upon independent selections, leading to joint probability of being included into the sample given straightforward by $\pi_{ij}=\pi_i \pi_j$ for any pair of units $i \neq j \in U$. This makes Poisson sampling especially friendly in the case of unequal probability selections when actual computation of joint inclusion probabilities can be an issue.
In practice, a (WOR) Poisson sampling design works as follows. Every population unit is {\em visited} sequentially and the decision whether to include or not include unit $i$ into the sample is made upon the result of independent Bernoulli trials each with probability of success equal to $\pi_i$; for instance, by randomly picking from an urn composed by the adequate number of coloured balls or by performing an equivalent computer experiment. 
As the simplest list-sequential design, Poisson sampling is given by the trivial choice for the updating rule;  that is, by choosing updating weights all equal to zero in equation (\ref{updating weights:def}). This essentially means no actual updating of probabilities in the updating matrix (\ref{updating matrix:def}),  those remaining are all equal to the initial ones $\pi_i^{(0)}=\pi_i$ for all population units at every step of the sequential selection.

\smallskip
PoSA proposal consists in integrating an adaptive component into a (plain) Poisson WOR sampling design.
For this basic proposal, the objective of the adaptive addition is the mere over-sampling of the positive cases which, according to the spatial pattern, can be expected {\em close} to each other along the chosen sequence.
With this purpose, the decision to include/not include unit $i$ in the sample is made depending on the previously observed value of the survey variable $y$. In terms of updating matrix, and more precisely of SMIs and updating rule, PoSA sampling can be described as follows. Given the set of initial probabilities $\pi_i^{(0)}, \ (i=1 \dots N)$, 
at the $1^{st}$ step of the selection $S_1$ has a Bernoulli distribution with initial probability $\pi_1^{(0)}$; at every subsequent step $i=2 \dots N$, $ S_i $ has Bernoulli distribution with probability $\pi_i^{(i-1)}$ given as a result of the updating at the previous step according to the following {\em adaptive} rule
\begin{equation}
\label{adaptive rule:PoSA} 
\pi_i^{(i-1)}=\left\{ \begin{array}{cl} 1 & \text{if } S_{i-1}=1 \ \text{and} \ y_{i-1}=1 \\ \pi_i^{(0)} & \text{otherwise}
\end{array} \right.
\end{equation}

Notice that, unlike the plain Poisson design, PoSA involves an actual step-by-step change of its updating matrix.
However, this updating is limited in that it actually affects only pairs of subsequent units, as defined {\em close} in the PoSA simplified setup. For all other (to be visited) units, the selection probability remains unaltered to their initial value. More precisely, at the $i$-th raw of the PoSA updating matrix, upon recording $S_i=s_i$, namely the selection/not selection of the visited unit $i$, every subsequent unit $j \ge i+1$ along such row will have selection probability $\pi_j^{(i)}$ updated as follows

\begin{equation}
\label{updating rule:PoSA}
\pi_j^{(i)}=\left\{ \begin{array}{cl} y_i s_i + \pi_{i+1}^{(0)}\left(1-y_i s_i\right) & \text{if} \quad j=i+1\\ \\
\pi_j^{(0)} & \text{if}\quad j > i+1 
\end{array} \right.
\end{equation}

\bigskip

As already mentioned, the updating matrix provides all the analytic input required for unbiased estimation. 
For PoSA SMI having Bernoulli distribution with updated probability given by (\ref{adaptive rule:PoSA}) according to (\ref{updating rule:PoSA}), it follows straightforward
\begin{equation}
\label{PoSA SMI:E}
E(S_i)=\pi_i^{(i-1)}= E\left(S_i^2\right) 
\end{equation}
and
\begin{equation}
\label{PoSA SMI:Var}
V(S_i)=E(S_i)[1-E(S_i)]=\pi_i^{(i-1)} \left(1-\pi_i^{(i-1)}\right)
\end{equation}

With regard to the selection of a pair of population units $j>i=1 \cdots N-1$ under PoSA sampling, we remark that the adaptive-updating rule only affects subsequent units: i.e., when $j=i+1$; otherwise, i.e., for all $j > i+1$, independence holds for PoSA being Poisson-based. Consequently, joint expectation for every pair of subsequent units is given by 
\begin{equation}
\label{PoSA SMI:mixed moment}
E\left(S_i S_{i+1}\right) = P\left(S_i=1, S_{i+1}=1\right) =P\left(S_i=1\right) P\left( S_{i+1}=1 \left | \right. S_i=1\right) = 
\pi_{i}^{(i-1)} \left[ y_i + \pi_{i+1}^{(0)}\left(1-y_i\right)\right] 
\end{equation}
leading to covariance
\begin{equation}
\label{PoSA SMI:cov}
Cov\left(S_i S_{i+1}\right) = \pi_{i}^{(i-1)} \left[ y_i + \pi_{i+1}^{(0)}\left(1-y_i\right) - \pi_{i+1}^{(i)}\right] 
\end{equation}
Otherwise $Cov\left(S_i S_{j}\right)=0$ for any pair of non-strictly subsequent units $j>i+1$. 
\smallskip

We now focus on the mean $\bar y= \sum_{i=1}^N y_i / N$ as population quantity of interest, which is in fact a proportion for binary $y$. This is, for instance, the case of our motivational example of TB prevalence surveys because prevalence is indeed defined as the proportion of TB cases in a given country at a given point in time.
Let $s$ denote the selected sample of population units, namely the set of $i$ such that the last $N^{th}$ row of the updating matrix (\ref{updating matrix:def}) shows a value $s_i$ equal to $1$. From equation (\ref{PoSA SMI:E}), an unbiased estimator is readily given 

\begin{equation}
\label{PoSA estimator}
\hat{\bar Y}_{PoSA}= \frac{1}{N}\sum_{i=1}^N y_i \frac{ S_i}{E(S_i)} = \frac{1}{N}\sum_{i=1}^N \frac{y_i S_i}{\pi_i^{(i-1)}} = \frac{1}{N}\sum_{i\in s} \frac{y_i}{\pi_i^{(i-1)}}
\end{equation}

For estimator (\ref{PoSA estimator}) being  of the HT type, its exact variance has the familiar closed form

$$V\left(\hat{\bar Y}_{PoSA}\right) = \frac{1}{N^2} \left[ \sum_{i=1}^N y_i^2 \frac{V\left(S_i\right)}{E\left(S_i\right)^2} + 2\sum_{i<j} \sum_{=2}^N y_i y_j \frac{Cov\left(S_i,S_j\right)}{E\left(S_i\right)E\left(S_j\right)}\right]$$
which significantly simplifies by excluding all null covariances, which happens between pairs of {\em distant } non-strictly subsequent units

$$V\left(\hat{\bar Y}_{PoSA}\right) = \frac{1}{N^2} \left[ \sum_{i=1}^N y_i^2 \frac{V\left(S_i\right)}{E\left(S_i\right)^2} + 2\sum_{i=1}^{N-1} y_i y_{i+1} \frac{Cov\left(S_i,S_{i+1}\right)}{E\left(S_i\right)E\left(S_{i+1}\right)}\right]$$
\begin{equation}
\label{PoSA Var}
= \frac{1}{N^2} \left[ \sum_{i=1}^N y_i^2 \ \frac{ 1-\pi_i^{(i-1)}}{\pi_i^{(i-1)}}+ 2\sum_{i=1}^{N-1} y_i y_{i+1} \frac{y_i + \pi_{i+1}^{(0)}\left(1-y_i\right) - \pi_{i+1}^{(i)}}{ \pi_{i+1}^{(i)}}\right].
\end{equation}

\smallskip
Finally, exact unbiased variance estimation follows in closed form again of the HT-type

\begin{equation}
\label{PoSA varest}
v\left(\hat{\bar Y}_{PoSA}\right) = \frac{1}{N^2} \left\{\sum_{i \in s} \left(\frac{y _i}{\pi_i^{(i-1)}}\right)^2 \left(1-\pi_i^{(i-1)}\right)+ 2\sum_{i \in s} y_i y_{i+1} \ \frac{y_i + \pi_{i+1}^{(0)}\left(1-y_i\right) - \pi_{i+1}^{(i)}}{\pi_{i+1}^{(i)}\pi_{i}^{(i-1)} \left[ y_i + \pi_{i+1}^{(0)}\left(1-y_i\right)\right] }\right\}
\end{equation}
where the right sum refers to selected units who were subsequent; {\em i.e., close} in the ordered population.

\smallskip
As we said earlier, PoSA was meant to be a starting point in a simplified setup. A basis from which, or even against which, to  develop improvements that are better tailored for the practical application at hand. With this in mind, we conclude the illustration of our PoSA proposal with two remarks. 

\smallskip
\noindent{\em Remark 1:}
for PoSA sampling being Poisson based, it results in a sample of random size. Random sample size is also a characteristic of adaptive sampling (see for instance Thompson \& Seber , 1996). However, random sample size is usually disliked by practitioners and professionals for making it difficult to plan in advance survey costs. A way to control the sample size is discussed in the next section. 

\smallskip
\noindent{\em Remark 2:} with regard to TB prevalence surveys, WHO guidelines currently suggest a multi-stage design under which selection units are in fact national sub-areas and all individuals in each selected area are invited for data collection. PoSA applies to this sort of set-up with a slight adjustment in the definition of its adaptive rule (\ref{adaptive rule:PoSA}).
Let $i$ indicate a group (sometimes called a {\em cluster}) of individuals (e.g. a national sub-area) and $N_i$ denote the size of $i$, so that the national population would have size $\sum_{i=1}^N N_i$. Let $y_i$ indicate the total number of positive cases detected in the selected area $i$. The adaptive rule can be given in terms of threshold $y_i/N_i > c_i \quad $ for a chosen level $c_i$, possibly different for different sub-areas. An example of such  threshold can be an anticipated guess or a previous estimate of the national prevalence in the country  to be used as cutoff. Thus, according to (\ref{adaptive rule:PoSA}), if in a selected area the prevalence is observed  that is greater than the  cutoff, then the {\em close} area (subsequent in the  sequence) should be certainly included in the sample, and otherwise selected with initial probability.

\bigskip\bigskip

\section{Controlling the Sample Size: Conditional Poisson Sequential Adaptive sampling}
\label{sec:4}
Preliminary simulations (Furfaro, 2017) have showed that PoSA has a notable  variability in the size of the final sample, with a tendency to provide both large samples, which directly affect survey costs, and small samples which can threaten estimation accuracy. This is also a characteristic of the plain Poisson sampling. To address the randomness around the sample size, a Conditional Poisson sampling, also known as Maximum Entropy, has been introduced in the literature where sample size is conditioned on  a chosen value (see for instance Till$\acute{\text e}$, 2006). With a similar approach, we now illustrate a Conditional version of PoSA, dubbed CPoSA, with the main purpose of gaining control over sample size at the design level of the survey. At the same time, we do not want to compromise on the desirable features of PoSA, namely the over-sampling of positive cases, its technical simplicity and readiness to implement. We therefore focus on the following key elements:

\begin{itemize}
\item[\em i)] To avoid unacceptable small sample, a {\em minimum} sample size should be established so that the selection process is not allowed to stop before this minimum has been reached; and
\item[\em ii)] To guarantee the over-sampling of positive cases, additional selections would be still allowed provided that they would result in additional positive cases.
\end{itemize}

In terms of SMIs and updating rule, the CPoSA algorithm is described as follows. Let $n_{min}$ be the pre-fixed required minimum sample size. Then, the set of initial probabilities has to be chosen such that $\sum_{i=1}^N \pi_i^{(0)}=n_{min}$. CPoSA selection is still list-sequential Poisson-based in that, similar to PoSA, at the $1^{st}$ step $S_1$ has a Bernoulli distribution with initial probability $\pi_1^{(0)}$ and at every subsequent step $i=2 \dots N$, $ S_i $ has Bernoulli distribution with updated probability $\pi_i^{(i-1)}$, located at the previous row, same column of the CPoSA updating matrix. However, unlike PoSA, the CPoSA updating rule is made both {\em adaptive}, according to the previous selection being or not a positive case, and also {\em dependent on} the number of units already visited, thus already recorded as either selected or non-selected in the sample. Moreover, to secure (at least) sample size $n_{min}$, the updating weights, as for the general updating rule (\ref{updating weights:def}), must have sum equal to $1$. A simple choice to accomplish this requirement is $w_{j-i}^{(i)}=1/(N-i),$ (Bondesson and Thorburn, 2008). This yields at the $i^{th}$ row of CPoSA updating matrix, for {\em all} selection probabilities of (to be visited) units $j \ge i+1$ to undergo an actual though constant updating as given by 

\begin{equation}
\label{updating rule:CPoSA}
\pi_j^{(i)}= \left\{\begin{array}{ll} 1 & \text{if} \quad j=i+1 \text{ and } s_i \cdot y_i=1 \\ 
\pi_j^{(i-1)} - \left(s_i - \pi_i^{(i-1)}\right)/(N-i) & \text{otherwise}
\end{array} \right.
\end{equation} 

\smallskip
\noindent It is important to remark that the CPoSA updating rule (\ref{updating rule:CPoSA}) depends exclusively on $s_i$; i.e., of the results of the $i^{th}$ selection for {\em all} units $j \ge i+1$. Hence, for non-strictly subsequent units, $j > i+1$ independence still holds, similarly to PoSA. As a result of CPoSA being in fact a slight modification of PoSA, in terms of SMIs and estimation mechanism, it can be modified quite straightforwardly.
In particular, equations (\ref{PoSA SMI:E}) and (\ref{PoSA SMI:Var}) still apply to CPoSA updating matrix, while the CPoSA joint expectation for strictly subsequent units’ results

\begin{equation}
\label{CPoSA SMI: mixed mom}
E\left(S_iS_{i+1}\right) = \pi_i^{(i-1)} \left\{ y_i + (1-y_i) \left[\pi_{i+1}^{(i-1)} - \left(1-\pi_{i+1}^{(i-1)}\right) (N-i)^{-1}\right]\right\}
\end{equation}
leading to covariance

\begin{equation}
\label{CPoSA SMI:cov}
Cov\left(S_i S_{i+1}\right) = \pi_i^{(i-1)} \left\{ y_i + (1-y_i) \left[\pi_{i+1}^{(i-1)} - \left(1-\pi_{i+1}^{(i-1)}\right) (N-i)^{-1} \right]- \pi_{i+1}^{(i)}\right\}
\end{equation}
Otherwise for pair of non-strictly subsequent unit independence holds leading to $Cov\left(S_i S_{j}\right)=0$ for all $j>i+1.$

With respect to estimation, equation (\ref{PoSA estimator}) as applied to CPoSA updating matrix still provides an unbiased estimator for the mean (proportion) of the (binary) survey variable $y$, with (exact) variance, simplified by excluding all zero covariances, given by

$$\hspace{-6cm}V\left(\hat{\bar Y}_{CPoSA}\right) = \frac{1}{N^2} \left[ \sum_{i=1}^N y_i^2 \ \frac{ 1-\pi_i^{(i-1)}}{\pi_i^{(i-1)}} \right. \ + $$
\smallskip
\begin{equation}
\label{CPoSA Var}
\left. + \ 2\sum_{i=1}^{N-1} y_i y_{i+1} \frac{y_i + (1-y_i) \left[\pi_{i+1}^{(i-1)} - \left(1-\pi_{i+1}^{(i-1)}\right) (N-i)^{-1} \right]- \pi_{i+1}^{(i)}}{ \pi_{i+1}^{(i)}}\right]
\end{equation}

We finally remark that, although CPoSA would certainly provide a sample size greater or equal to the chosen $n_{min}$, the final sample will possibly contain additional positive cases, if detected, so that the actual sample size would be in fact random. Consequently, similarly to PoSA, CPoSA unbiased variance estimation is given in closed form of the HT type

$$\hspace{-6cm} v\left(\hat{\bar Y}_{CPoSA}\right) = \frac{1}{N^2} \left\{\sum_{i \in s} \left(\frac{y _i}{\pi_i^{(i-1)}}\right)^2 \left(1-\pi_i^{(i-1)}\right) \right. \ +$$
\begin{equation}
\label{CPoSA varest}
\left. + \ 2\sum_{i \in s} y_i y_{i+1} \ \frac{y_i + (1-y_i) \left[\pi_{i+1}^{(i-1)} - \left(1-\pi_{i+1}^{(i-1)}\right) (N-i)^{-1} \right]- \pi_{i+1}^{(i)}}{ \pi_{i+1}^{(i)} \pi_i^{(i-1)} \left\{ y_i + (1-y_i) \left[\pi_{i+1}^{(i-1)} - \left(1-\pi_{i+1}^{(i-1)}\right) (N-i)^{-1}\right]\right\}}\right\}
\end{equation}
where, again, in the right sum only the selected units who were strictly subsequent in the ordered population are concerned.

\bigskip \bigskip

\section{Some Empirical Evidence}
\label{sec:5}
Our motivational example also offers a rich source of real data  and a natural basis for  discussing  the application potential of the innovative sampling strategies proposed in this paper. 

In 2007, the WHO Global Task Force on TB Impact Measurement identified a set of 22 global focus countries, which  account for about 80\% of the world TB burden,  to receive priority attention and support.  As an example, Table \ref{tab:realdata} reports data from 18 TB prevalence surveys conducted in Asia in the period 1990--2012 (Onozaki et al., 2015).  Most of those surveys were operated  with the technical support of WHO, according to standardised guidelines  for survey design, implementation, analysis and reporting. All of the participants  underwent both symptom screening and chest X-ray,  and positive cases were defined both as bacteriologically confirmed via a laboratory test ($Bact^+$) and as smear positive via a sputum specimen on-site examination ($S^+$).  

Table \ref{tab:realdata} highlights what we pointed out as limits of applying a traditional sampling design for collecting data of a rare disease: to attain a {\em representative sample}, very large sample sizes are required  for an acceptable estimation precision  (recommended in the range  20 to 25\%) versus a  small amount of  positive cases detected  for treatment,  and this consequently inflated the cost-per-case detected. In fact  the  case detection rate per 100 participants enrolled in the survey ({\em i.e.} 100 times the ratio of  positive cases over participants, as displayed in boldface in parentheses) ranges from 0.14\% to 1.22\%  for bacteriologically confirmed TB cases and even less  for smear positive TB cases (i.e. 0.06\% to 0.36\%), the latter being the most infectious condition.

The potential to over-sample positive-cases is apparent, pursued by the adaptive component of PoSA and CPoSA  for correcting   such effects.  
A more recent example can be found in the data  from the Kenya tuberculosis prevalence survey  (Ministry of Health, Republic of Kenia,  2016), which was  collected nationwide  cross-sectionally   between July 2015 and July 2016. 
A total of 305  positive  cases were diagnosed (all forms pulmonary TB) among 63,050 participants. The overall  survey budget was  more than 5 million USD (Annex 1), resulting in cost-per-case-detected greater than 16,300 USD.  Logistics  accounted for the largest part of these costs, with  $38.4\%$ for cluster budget,  including the organisation and management of primary sampling units, and  another $9\%$ for transport.  This suggests the potential  of the sequential component of  PoSa and CPoSA, which is intended as a flexible and more efficient environment  for  dealing with logistical and budget constraints.


\begin{table} \centering
$$\begin{array}{lcccc}
	&	n	& S^+		&	Bact^+		\\
\hline									
\text{Bangladesh 2008}	&	52098	&	33	({\bf 0.06})	&	\text{not comparable}	\\
\text{Cambodia 2002}	&	22160	&	81	({\bf 0.37})	&	271	({\bf  1.22 })	\\
\text{Cambodia 2011}	&	37417	&	103	({\bf 0.28})	&	314	({\bf  0.84 })	 \\
\text{China 1990}	&	1461190	&	1827	({\bf 0.13})	&	2389	({\bf  0.16 })	\\
\text{China 2000}	&	365097	&	447	({\bf 0.12})	&	584	({\bf  0.16 })	\\
\text{China 2010}	&	252940	&	188	({\bf 0.07})	&	347	({\bf  0.14 })	\\
I\text{Indonesia 2004}	&	50154	&	80	({\bf 0.16})	&	\text{not done}		\\
\text{Lao PDR 2011}	&	39212	&	107	({\bf 0.27})	&	237	({\bf  0.60 })	\\
\text{Myanmar 1994}	&	37424	&	39	({\bf 0.10})	&	\text{not done}		\\
\text{Myanmar 2009}	&	51367	&	123	({\bf 0.24})	&	311	({\bf  0.61 })	\\
\text{Pakistan 2011}	&	105913	&	233	({\bf 0.22})	&	314	({\bf  0.30 })	\\
\text{Philippines 1997}	&	12850	&	47	({\bf 0.37})	&	127	({\bf  0.99 })	\\
\text{Philippines 2007}	&	20625	&	60	({\bf 0,.29})	&	151	({\bf  0.73 })	\\
\text{Republic of Korea 1990}	&	48976	&	70	({\bf 0.14})	&	118	({\bf  0.24 })	\\
\text{Republic of Korea 1995}	&	64713	&	60	({\bf 0.09})	&	142	({\bf  0.22 })	\\
\text{Thailand 1991}	&	35844	&	73	({\bf 0.20})	&	101	({\bf  0.28 })	\\
\text{Thailand 2012}	&	62536	&	58	({\bf 0.09})	&	142	({\bf  0.23 })	\\
\text{Viet Nam 2007	} &	94179	&	174	({\bf 0.18})	&	269	({\bf  0.29 })	 \\
\hline									
\end{array}$$
\caption{\small{Number of participants  and number of positive cases  observed  ({\bf \% case detection rate}) for 18 TB surveys, Asia 1990-2012 }}
\label{tab:realdata}
\end{table}

We now illustrate a simulation study  that aims to  explore the strengths and weaknesses  of PoSA and CPoSA  against a traditional (cross-sectional, non-adaptive)  sampling design. The simulation setup is a simplification of the design suggested in the WHO guidelines  for TB prevalence surveys (see WHO, 2011;  Chapter 5). The country area to be surveyed is divided into a chosen number $M$ of sub-areas, informed in such a way that each includes a recommended equal/similar number of population units. A pre-defined number $m$ of sub-areas is selected at random and all individuals living in the selected area are invited to participate in the survey. The size $m$ of such first-stage sample of areas is computed according to the WHO guidelines (see WHO, 2011 Chapter 5, formula 5.4) for the recommended level of estimate precision ($25\%$) and based on a preliminary guess of the national TB prevalence; i.e., the population quantity to be estimated. Notice that the true population prevalence has been used as {\em guess} in the computation of $m$ for simulation purposes, so that the traditional benchmark design has been simulated in its {\em best} scenario.

The final sample size $n$ is then given by the total number of individuals included into the $m$ selected areas. Further note that the suggested computation of $m$ is also function of a quantity $k$, which adjusts for between-area variability; namely, {\em how different} might each area-prevalence be compared with the overall national prevalence, ultimately resulting in the “design effect”. $k$ is defined as a {\em coefficient of between-area variation} and it is usually informed based on previous surveys or expert opinion. In more detail, $k$ is the ratio of the between-area standard deviation of the binary study variable $y$ (positive or negative TB case) over the population mean $\bar Y$; i.e., the true national prevalence.

Previous empirical evidence from an extended simulation study (Furfaro, 2017) has identified $k$ as a key simulation factor, the most influential in comparing the performance of the three sampling strategies. An equally important simulation specification is the spatial clustering of the study variable, namely how concentrated or else spread-out are TB cases over the surveyed region. Indeed, the greater the tendency of positive cases to gather in particular sub-areas, from now on as referred as {\em clusters} of (positive) cases, the greater the between area variability as measured by $k$.

\begin{figure}[!h]
\centering
\includegraphics[width=\textwidth]{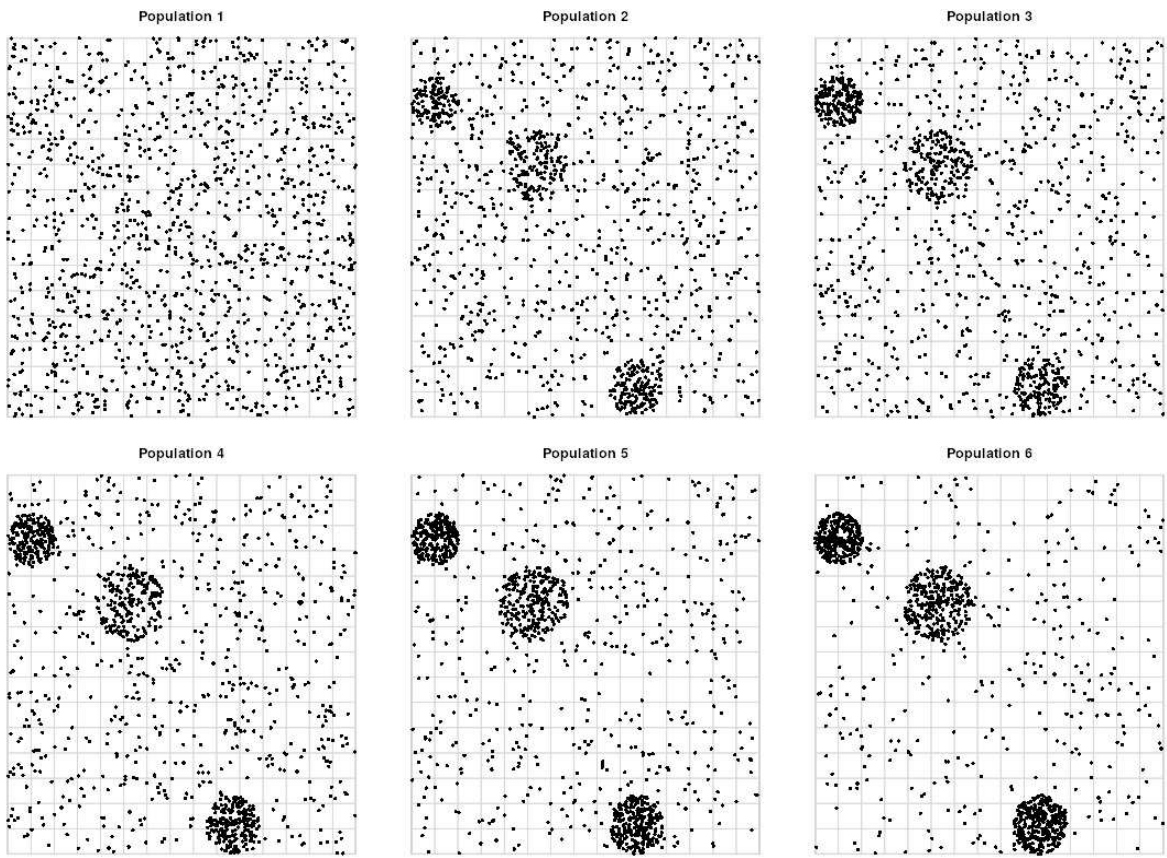} 
\caption{\small{Six simulated populations: dots depict positive cases.}}
\label{fig:pops}
\end{figure}

The population size $N$, the number and the shape of the clusters for a given level of $k$, and the actual value of the national prevalence to be estimated appear to not be influential with respect to the relative performance pattern of the strategies under comparison. 
Thus the simulation scenario comprises $6$ populations, each of size $N= 250000$ with a $0.5\%$ national prevalence $(\bar Y=0.005)$ and increasing proportion of positive cases gathered into $3$ clusters, leading to increasing levels of between-areas variation $k$, as described in Table \ref{tab:pops}. 
The six simulated populations are depicted in Figure \ref{fig:pops} as partitioned in $M=225$ sub-areas (via the super-imposed 15x15 grid). These sub-areas have been used as primary sampling units under each of the three simulated strategies and all population units included into the selected areas have been included in the final sample. For all three strategies, the same set of equal area-selection probabilities has been applied.

\begin{table} \centering

$$\begin{array}{c|cccccc}
\text{\small Pop} & 1 & 2 & 3 & 4 & 5 & 6 \\
\hline
\% \text{ \small of cases} & & & & & & \\
\text{\small gathered into} & 0\% & 30\% & 40\% & 47\% & 60\% & 70\% \\
\text{\small the 3 clusters} & & & & & & \\
\hline
k & 0.5 & 1.1 & 1.4 & 1.7 & 2.0, & 2.5 \\
\end{array}$$
\caption{\small{ Key features of the six simulated populations }}
\label{tab:pops}
\end{table}

Every simulation is based on 5000 Monte Carlo (MC) runs. 
The following four key aspects have been explored for each of the three simulated sampling strategies: 
\begin{enumerate}
\item The final sample size $n$, which is fixed for the traditional benchmark design while it is random for both PoSA and CPoSA designs.
\item The accuracy of the final estimate, as measured by the MC Root Mean Squared Error of the estimator, under a given sampling design: 
$\sqrt{E_{MC} \left(\hat{\bar Y} - \bar Y \right)^2 }$.
\item The ability to detect positive cases, as measured by the MC Expectation of the rate of positive cases into all simulated sample, under a given design: $E_{MC} \left( \sum_{k= 1}^n y_k / n \right)$.
\item The cost per (positive) case detected, as measured by the MC Expectation of the ratio of the total survey cost over the number of (positive) cases selected in every simulated sample $E_{MC} \left( C/ \sum_{k = 1}^n y_k\right)$. The survey cost has been computed under a conventional linear cost function 

\begin{equation}
C = c_0+ c_1 m+\sum_{j=1}^m N_j c_2 = 100000 + 1000 m+\sum_{j=1}^m 10 N_j 
\label{eq: costs}
\end{equation}
where $c_0$ denotes the fixed cost while $c_1$ and $c_2$ are, respectively, the unitary cost for every selected area (containing $N_j$ individuals) and the unitary cost for collecting data at every unit included in the selected area. Limited to PoSA and CPoSA, a fixed discount $(20\%)$ has been applied to $C$, with the purpose of simulating the expected savings following from the increased control options over logistics and budget, such as  the possibility to plan the route for sequential sample selection by minimising travel costs. 
\end{enumerate}

\setcounter{figure}{1}
\begin{figure}[!h]
\centering
\includegraphics[width=\textwidth]{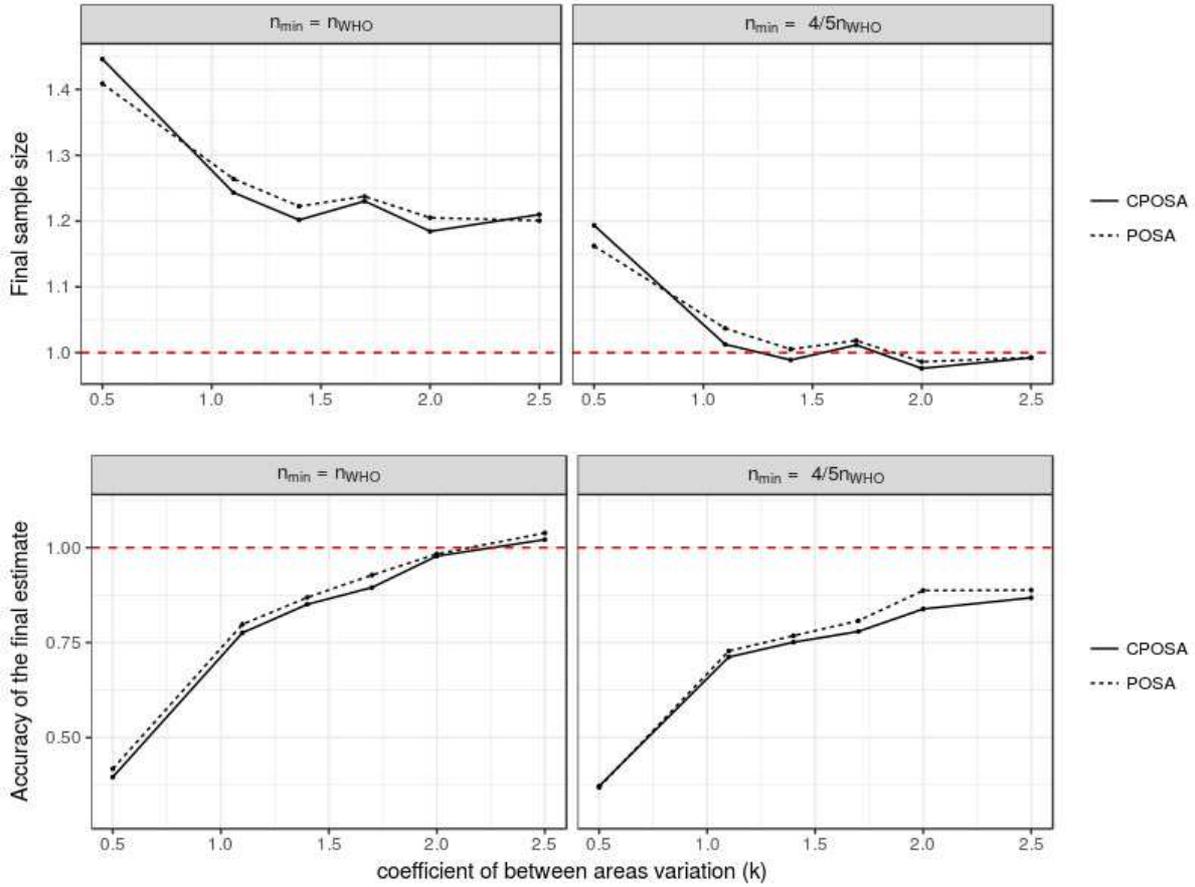} 
\caption{\small{Final sample size (above) and accuracy of the final estimate (below): ratio over the traditional benchmark design (dashed line means equal performance) for increasing level of $k$ }}
\label{fig:mcperformance1_bis}
\end{figure}

Simulation results for the MC measures of performance 1 to 4 above, are presented in Figures \ref{fig:mcperformance1_bis} and \ref{fig:mcperformance2} where the left-hand panels refer to a minimum sample size for CPoSA set equal to the sample size computed according to WHO guidelines for the traditional benchmark design $(n_{min}= n_{WHO})$ while the right-hand panels refer to $n_{min}$ for CPoSA slightly less than the WHO sample size $(n_{min}=\frac{4}{5} n_{WHO})$.
To  facilitate comparisons,  the results are presented for PoSA and CPoSA as a ratio {\em relative} to the traditional benchmark design. Therefore, the graphs depict a dashed line (corresponding to 1) which would indicate equal performance of the proposed strategy with respect to the traditional benchmark, while gains/losses show otherwise above/ below the dashed equality line.

The empirical results confirm between-area variability (as measured by $k$) as the driving factor for comparing a sequential-adaptive strategy over a traditional strategy. Both PoSA and CPoSA show uniformly improved performances as the between-areas variability increases, moderate to high intra-area variability/correlation being naturally the case for rare and spatially clustered populations. In this sense, the proposed strategies show their capacity to exploit and benefit on the existing uneven spread/concentration of positive cases across the surveyed region. Meanwhile, the traditional benchmark design appears to undergo intra-area variation, leading to the need to deal with it; for example, by forcing sub-areas of equal size and/or requiring controlling between-area variation that should be kept in a recommended low range (see for instance  WHO, 2011 Chapter 5).

Figure \ref{fig:mcperformance1_bis} highlights that both PoSA and CPoSA may lead to an increased final sample size, up to $1.5$ times the fixed sample size of the traditional benchmark design (top left-hand panel). However, such effect rapidly decreases and stabilises as the between-area variability increases. Also notice that under CPoSA, it is guaranteed that possibly additional sampled units would be positive cases. Nevertheless, unplanned large sample sizes, usually an undesirable practical issue, can be managed or even avoided by setting smaller values for $n_{min}$, as shown by the top right-hand panel in Figure \ref{fig:mcperformance1_bis}. Similarly, the accuracy loss of the final estimate showed by the bottom panels of Figure \ref{fig:mcperformance1_bis} for low values of $k$, rapidly reduces as $k$ increases, reaching the same accuracy provided by the traditional design for larger sample sizes (bottom left-hand panel) and stabilising around a $10\%$ loss for equal sample sizes (bottom right-hand panel).

\setcounter{figure}{2}
\begin{figure}[!h]
\centering
\includegraphics[width=\textwidth]{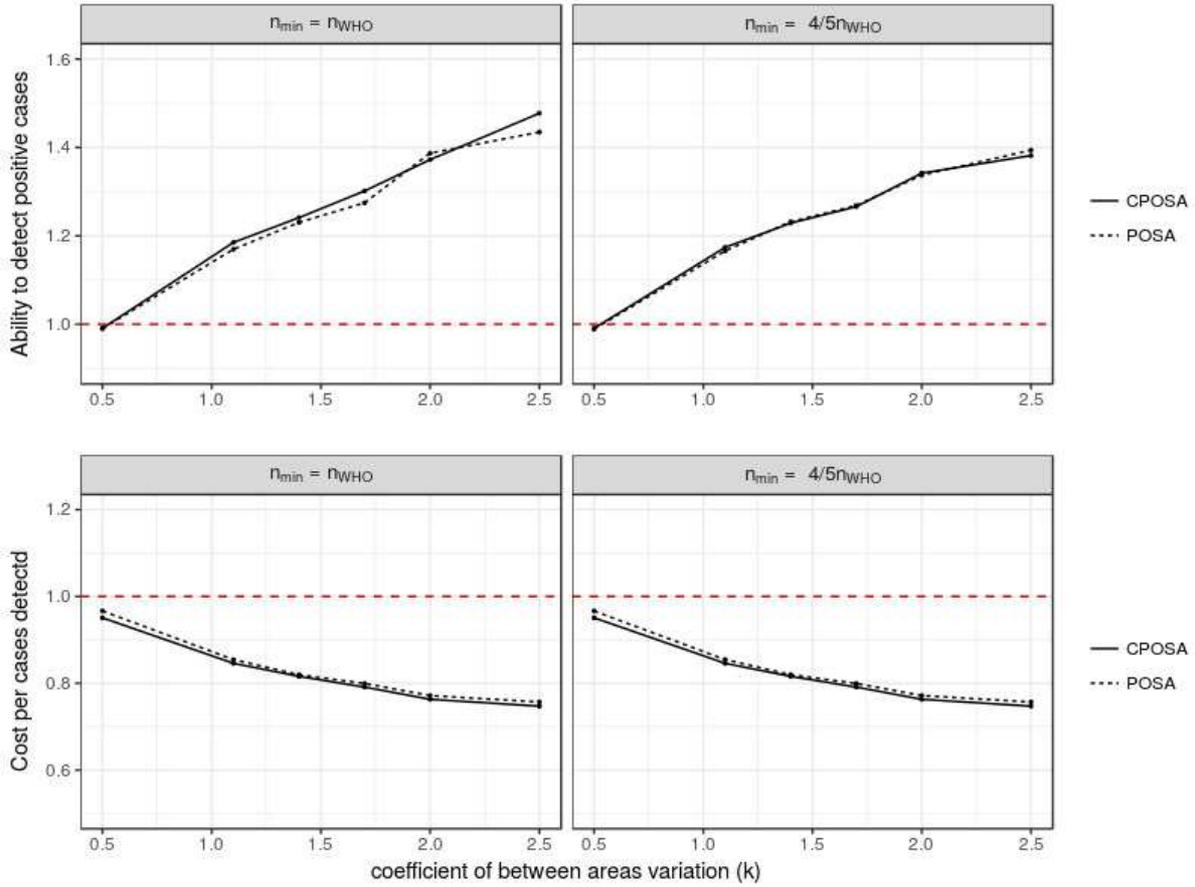} 
\caption{\small{Ability to detect cases (above) and cost per case detected (below): ratio over the traditional benchmark design (dashed line means equal performance) for increasing level of $k$ }}
\label{fig:mcperformance2}
\end{figure}

The possibly larger sample size and smaller estimate accuracy somehow seem to be the expected price to pay when the over-sampling of positive cases is a strategic main goal of the study at hand. Figure \ref{fig:mcperformance2} provides empirical evidence of the advantages offered by the proposed strategies over the traditional benchmark. The enhanced ability of over-detecting cases clearly shown by the proposed strategies (top panels) leads to downsized cost-per-case-detected, and such effect uniformly and quickly increases as the between-area variability increases (bottom panels). Under a sequential adaptive strategy the samples may include up to $1.5$ times positive cases than under a traditional design with up to a $25\%$ reduction of cost-per-case detected.

\bigskip \bigskip
\section{Concluding Remarks and Research Perspectives}

In this paper, an innovative sampling strategy is proposed, which applies to population-based surveys for a rare and clustered trait. The underlying idea is to integrate an adaptive component into a sequential selection, with the aim of simultaneously enhancing the detection of positive cases by exploiting the spatial pattern and offering a flexible framework for managing logistical constraints at the design level of the survey. Moreover, a sequential step-by-step selection process would naturally allow for real-time adjustments of the data collection during field operations. An increased effectiveness of the large budget usually required by observational studies for rare diseases, is expected as a result.

A sequential adaptive Poisson-based PoSA sampling design and unbiased estimation have been presented. A conditional version, CPoSA,  has also been introduced to increase the control over the size of the final sample (namely, guaranteeing a pre-chosen minimum sample size) and allow for the additional selection of any further positive cases that are adaptively detected.

Empirical evidence, via Monte Carlo simulations, has been presented, which makes the case for PoSA and CPoSA as recommendable choices over traditional, multi-stage, cross-sectional, self-weighted sampling designs, for moderate to high between-area variability and spatial clustering. Indeed, the greater the tendency of positive cases to gather in particular areas of the surveyed region (e.g. hot-spots of an infectious disease), the greater the case-detection ability of the proposed strategies are, which leads to a fall in the cost-per-case detected.

Given that PoSA and CPoSA are developed as first proposals in a simplified set-up, some limitations have also been demonstrated, which in fact open future research opportunities.

First, we remark that both PoSA and CPoSA are in fact non-fixed sample size designs. This feature is typical of adaptive sampling; however, a trade-off may exist between the goal of oversampling positive cases and crucial requisites for a careful and in-advance planning of the total survey cost. In real applications, additional sampling selections might not be allowed, even if they were additional positive cases, which is assured with the proposed CPoSA. A methodological advancement to PoSA should therefore consider the choice of a maximum sample size $n_{max}$, as established either by its own or jointly with a chosen $n_{min}$, so that the sampling selection must stop if such maximum size is reached. Because this would happen at any step $j \ge n_{max}, $ the selection sequence could be trimmed before all units are visited, which makes unbiased estimation more complicated than either the PoSA or CPoSA estimation. More research is needed to deal with such a conflict with {\em simplicity}; i.e., the main objective of the fostered improved strategy.

Second, a natural improvement would be to relax the assumption of a linearly ordered population (e.g. the pre-fixed path in the TB example) to  allow a two-dimensional sequential selection that is able to move freely all along the geographical area of interest. This could be done by leaving the Poisson list-sequential choice in favour of a {\em Spatially Correlated Poisson Sample}  (Grafstr$\ddot{\text o}$m, 2012), which is a recent ready-to-implement proposal generalising the original {\em Correlated Poisson Sample} (Bondesson \& Thorburn, 2008). This is rooted in the idea of controlling covariances between SMIs of population units, either {\em close} or {\em distant} in space, by means of a feasible choice of the updating weights, to obtain a {\em spatial balance}; namely, a selected sample well diffused upon the region of interest. 

Future research will explore two further directions: the availability of auxiliary information at the design level of the survey and the production of disaggregated estimates. Auxiliary variable and accessible meta data could be effectively employed both at the design stage and at the estimation  stage of the survey. For instance, in our motivational TB example, epidemiological socio-cultural and/or economic meta data may be available from previous surveys and official registers. They can be used for an advanced definition of neighbourhood conditions in the space, thus refining the mere geographical proximity applied to PoSA and CPoSA. A refitted definition of close/distant units should also foster a boosting effect upon the ability of the sampling design to detect positive cases, consequently protecting against the tendency of PoSA to produce also small and very small samples. As a matter of fact, both PoSA and CPoSA over-sampling depends on the detection of a first positive case included in a hot-spot area, which would ignite the adaptive rule at work.

The availability of auxiliary information will also allow for advanced estimation beyond HT type Inverse-probability-weighting, toward regression and calibration methodology, with potential for significant gains in the accuracy of the released estimates, see  S$\ddot{\text a}$rndal  {\em et al.} (1992)  for a general introduction to model-assisted estimation. 

Finally {\em Small Area Estimation}  (SAE) will be considered, in the framework of Sequential Adaptive sampling, which is proposed here, with the purpose of producing disaggregated estimates, see Rao \& Molina  (2015) for a comprehensive account to SEA. This is motivated again by TB prevalence surveys: a crucial need exists for sub-national estimates, as testified by the increasingly strong request from countries, funding and international health agencies. 

\bigskip\bigskip
\begin{singlespace}
\noindent{\bf References} \\

{\small  
\begin{hangparas}{.25in}{1}

Bondesson, L. \& Thorburn, D. (2008). A list-sequential sampling method suitable for real-time sampling. {\em Scand. J. of Statistics},  {\bf 35}, 466--483 

Floyd , S., Sismanidis, C., Mecatti, F., Floyd , K.  \&Yamada N. (2011). {\em Sampling design}. In {\em Tuberculosis Prevalence Surveys: a handbook} , Chapter 5, 51-80. Geneva: World Health Organization ({\em https://www.who.int/tb/advisory\_bodies/impact\_measurement\_taskforce/resources\_documents/thelimebook/en/}

Furfaro, E. (2017). {\em A sequential adaptive approach for surveying rare and clustered populations}. PhD Thesis, University of Milano-Bicocca ({\em to appear online at} https://boa.unimib.it)

Grafstr$\ddot{\text o}$m,  A. (2012). Spatially correlated Poisson sampling. {\em Journal of Statistical Planning and Inference} {\bf 142}, 139--147

H$\acute{\text{a}}$jek,  J. (1964). Asymptotic theory of rejective sampling with varying probabilities from a finite population. {\em The Annals of Mathematical Statistics}, {\bf 14}, 333--362

Hansen,  M.H. \& Hurwitz, W.N. (1943). On the Theory of Sampling from Finite Populations. {\em The Annals of Mathematical Statistics}, {\bf 14}, 333--362

Horvitz, D.G. \& Thompson,  D.J. (1952). A Generalization of Sampling Without Replacement from a Finite Universe, {\em Journal of the American Statistical Association}, {\em 47}, 663--685

WHO (2011). {\em Tuberculosis prevalence surveys: a handbook} (The Lime Book). Geneva: World Health Organization 
http://www.who.int/tb/advisory$\text{\_}$bodies/impact$\text{\_}$measurement$\text{\_}$taskforce/resources$\text{\_}$documents/thelimebook/en

JOS Special Issue on Adaptive Designs (2017). {\em Journal of Official Statistics}, {\bf 33}, http://www.scb.se/en

Kalton, G. \& Anderson, D.W. (1986). Sampling Rare Populations
{\em Journal of the Royal Statistical Society. Series A}, {\bf 149}, 65--82

Ministry of Health, Republic of Kenia  (2016). {\em Kenya Tubercolosis Prevalence Survey: Final Survey report}.  https://www.chskenya.org/wp-content/uploads/2018/04/Final-TB-Prevalence-Survey-Report.pdf

Onozaki, I.,  Law, I.,  Sismanidis, C.,  Zignol, M., Glaziou, P. \& Floyd, K.  (2015). National tuberculosis prevalence surveys in Asia, 1990-2012: an overview of  results and lessons learned. {\em Tropical Medicine and  International Health}, {\bf 20},  1128--1145 

Rao, J.N.K \& Molina, I. (2015). {\em Small Area Estimation.} 2nd ed. New Jersey: Wiley

Seber, G.A.F. \& Salehi , M.M. (2012). {\em Adaptive Sampling Designs}. New York: Springer

Sirken, M.G. (2004). Network sample surveys of rare and elusive population: a historical review. {\em Proceedings of Statistics Canada Symposium 2004:} Innovative Methods for Surveying Difficult-to-reach Populations. [CD-RM]

Thompson, S.K. (1990). Adaptive Cluster Sampling. {\em Journal of the American Statistical Association}, {\bf 85}, 1050--1059 

Thompson, S.K. (2006). Adaptive Web Sampling. {\em Biometrics} {\bf 62}, 1224--1234

Thompson, S.K. (2017). Adaptive and Network Sampling for Inference and Interventions in Changing Populations. {\em Journal of Survey Statistics and Methodology}, {\bf 5}, 1--21

Thompson, S.K \& Seber, G.A.F (1996). {\em Adaptive Sampling}. Wiley Series in Probability and Statistics, New Jersey: Wiley

Till$\acute{\text e,}$ Y. (2006). {\em Sampling Algorithms}. New York: Springer

 Tourangeau, R., Brick, J.M., Lohr, S. \& Li J. (2017). Adaptive and responsive survey designs: a review and assessment. {\em Journal of the Royal Statistical Society, Series A}. {\bf 180}, 203--223

S$\ddot{\text a}$rndal, C.E., Swensson, B. \&  Wretman, J. (1992). {\em Model Assisted Survey Sampling}. New York: Springer-Verlag

WHO (2018). {\em Global Tuberculosis Report } $https://www.who.int/tb/publications/global_report/en/$

\end{hangparas}
} 
\end{singlespace}
\end{document}